\newif\ifpdf
\ifx\pdfoutput\undefined
\pdffalse 
\else
\pdfoutput = 1 
\pdftrue
\fi

\documentclass [11pt, twocolumn, a4paper, openright, oneside] {article}

\usepackage {amsmath, amssymb, amsfonts}
\usepackage {graphics}
\usepackage {graphicx}
\usepackage {longtable}
\usepackage {pgf,tikz}
\usepackage {float,afterpage} 
\usepackage {verbatim} 
\usetikzlibrary{arrows,automata,shapes,snakes,backgrounds,topaths,petri}
\usepackage {subfigure}

\title{Tutorial on ABC rejection and ABC SMC for parameter estimation and model selection}
\date{}
\author{Tina Toni, Michael P. H. Stumpf\\ \footnotesize{ttoni@imperial.ac.uk, m.stumpf@imperial.ac.uk}}

\begin{document}

\onecolumn

\maketitle

In this tutorial we schematically illustrate four algorithms:

\begin{enumerate}
 \item ABC rejection for parameter estimation \cite{Beaumont:2002p13862, Marjoram:2003p5},
 \item ABC SMC for parameter estimation \cite{Sisson:2007p2, Toni:2009p20998, Beaumont:2009p30650},
 \item ABC rejection for model selection on the joint space \cite{Grelaud:2009p29539},
 \item ABC SMC for model selection on the joint space \cite{Toni:2010p30719}.
\end{enumerate}

We suggest to read this tutorial from the beginning. We start with a detailed explanation of the ABC rejection algorithm, which later helps to understand ABC SMC as it is based on the same concepts. Also, both model selection algorithms are closely related to parameter estimation algorithms and it is therefore helpful to understand those first.\\
\\
This tutorial forms a part of the supplementary material of the paper "Simulation-based model selection for dynamical systems in systems and population biology, Bioinformatics, 26 (1), 104-110,  2010" (T. Toni, M. P. H. Stumpf).

\newpage

\twocolumn

\begin{figure}[thp]	
	\centering

	\subfigure[]{\includegraphics[width=2.5cm]{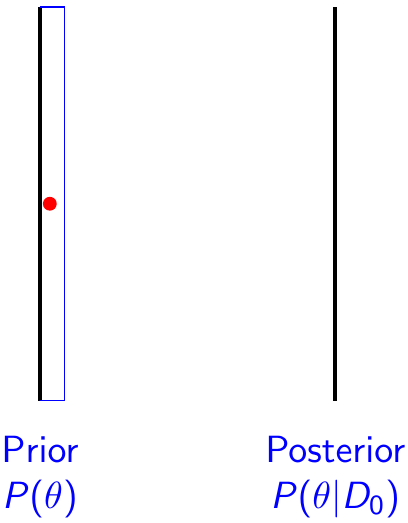}\label{fig:1}} \\
	\subfigure[]{\includegraphics[width=2.5cm]{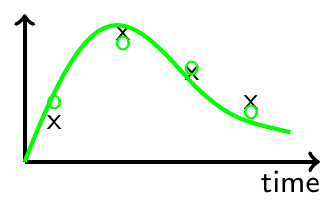}\label{fig:2}} \\
	\subfigure[]{\includegraphics[width=2.5cm]{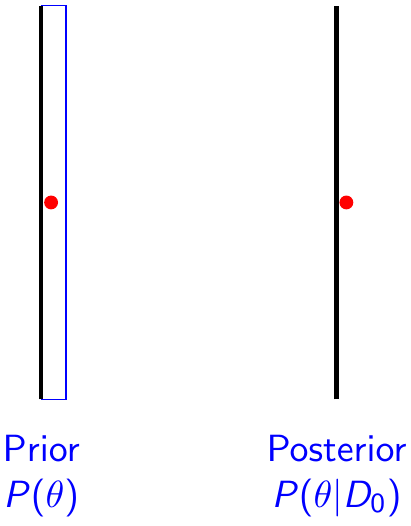}\label{fig:3}} \\
	\subfigure[]{\includegraphics[width=5.5cm]{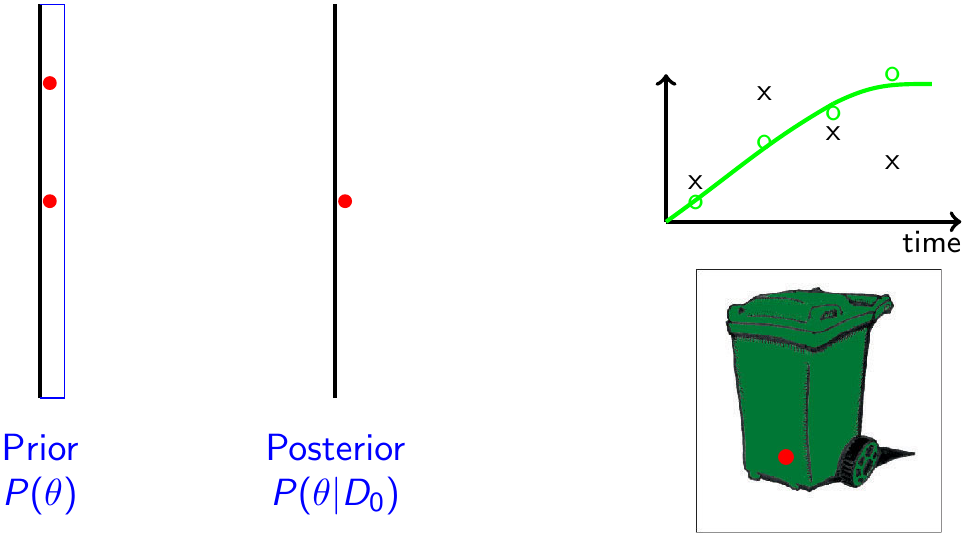}\label{fig:4}} \\	
	\subfigure[]{\includegraphics[width=2.5cm]{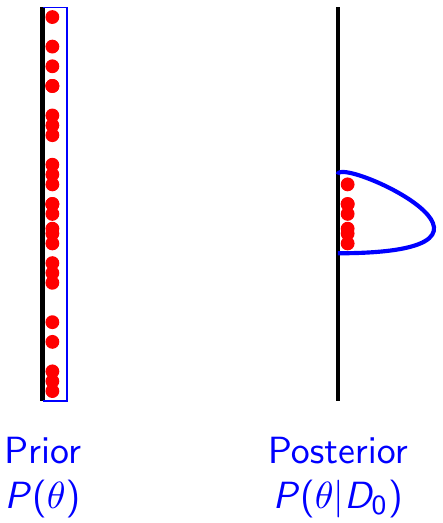}\label{fig:5a}} \hspace{1cm}
	\subfigure[]{\includegraphics[width=2cm]{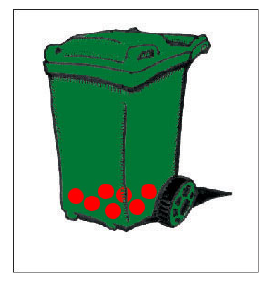}\label{fig:5b}}		
	
			\caption{\small{Schematic representation of ABC rejection.}}

\end{figure}
\small
\textbf{ABC rejection}\\
\\
(a) We define a prior distribution $P(\theta)$ and we would like to approximate the posterior distribution $P(\theta|D_0)$. We start by sampling a parameter $\theta^{*}$ from the prior distribution. We call this sampled parameter \textit{a particle}.\\
\\
(b) We simulate a data set $D^*$ according to some simulation framework $f(D|\theta^{*})$. In our examples we use different simulation frameworks. If we simulate a deterministic dynamical model, we add some noise at the time points of interest. If we simulate a stochastic dynamical model, we do not add any additional noise to the trajectories. We compare the simulated data set $D^*$ (circles) to the experimental data $D_0$ (crosses) using a distance function, $d$, and tolerance $\epsilon$; if $d(D_0, D^*) \leq \epsilon$, we accept $\theta^*$. The tolerance $\epsilon \geq 0$ is the desired level of agreement between  $D_0$ and $D^*$. \\
\\	
(c) The particle $\theta^{*}$ is accepted because $D^{*}$ and $D_0$ are sufficiently close.\\
\\
(d) We sample another parameter $\theta^{*}$ from the prior distribution and simulate a corresponding dataset $D^{*}$. In this case $D^{*}$ and $D_0$ are very different and we reject the particle (we "throw it away").\\
\\
(e) We repeat the whole procedure until N particles have been accepted. They represent a sample from $P(\theta|d(D_0,D^{*})\leq \epsilon)$, which approximates the posterior distribution. If $\epsilon$ is sufficiently small then the distribution $P(\theta|d(D_0, D^*) \leq \epsilon)$ will be a good approximation for the ``true'' posterior distribution, $P(\theta|D_0)$.\\
\\
(f) Many particles were rejected in the procedure, for which we have spent a lot of computational effort for simulation. ABC rejection is therefore computationally inefficient. We can use ABC SMC to reduce the computational cost. 

\newpage

\begin{figure}[thp]	
	\centering

	\subfigure[]{\includegraphics[width=7.2cm]{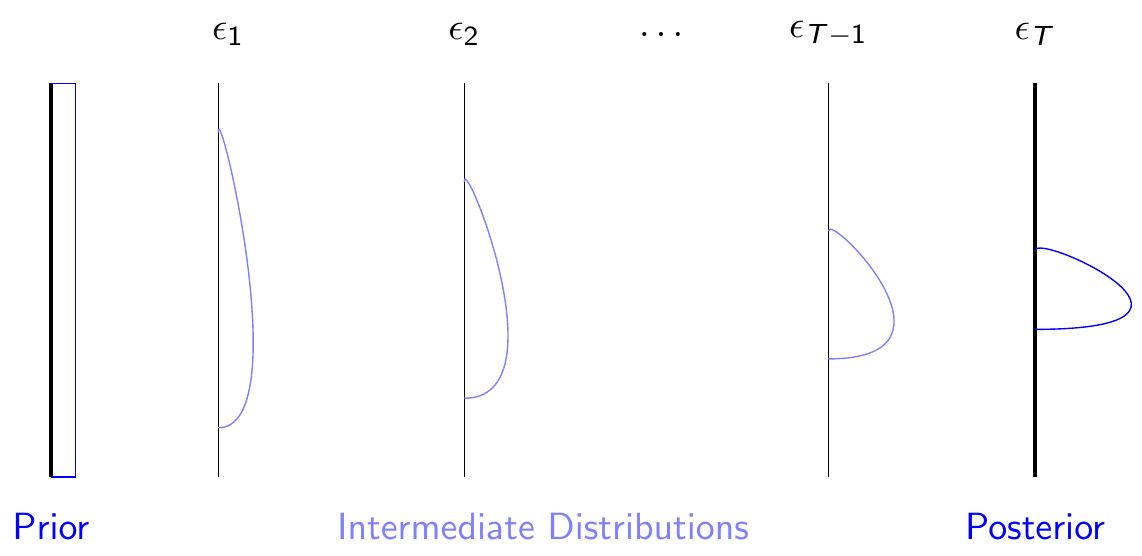}\label{fig:1smc}} \\
	\vspace{0.5cm}
	\subfigure[]{\includegraphics[width=7.2cm]{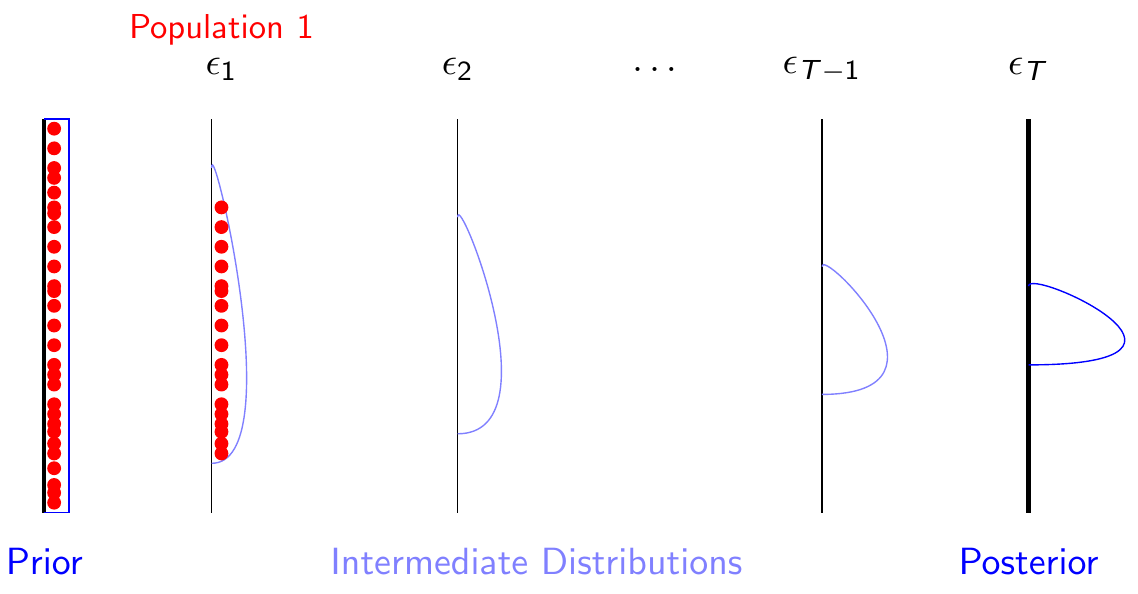}\label{fig:2smc}} \\
	\vspace{0.5cm}
	\subfigure[]{\includegraphics[width=7.2cm]{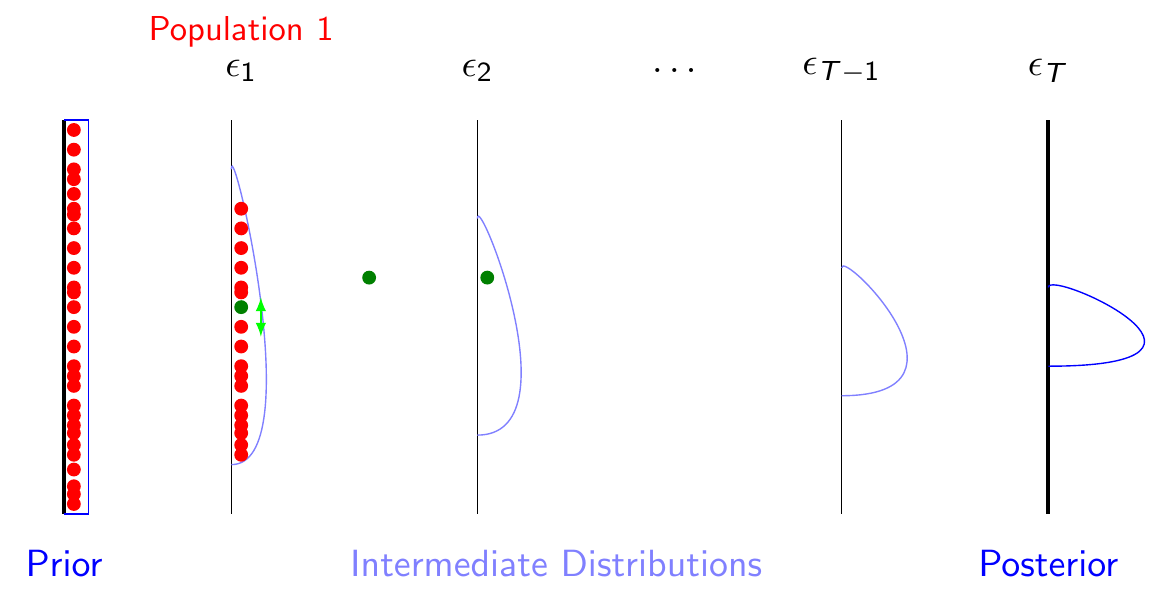}\label{fig:3smc}} \\
	\vspace{0.5cm} 
	\subfigure[]{\includegraphics[width=7.2cm]{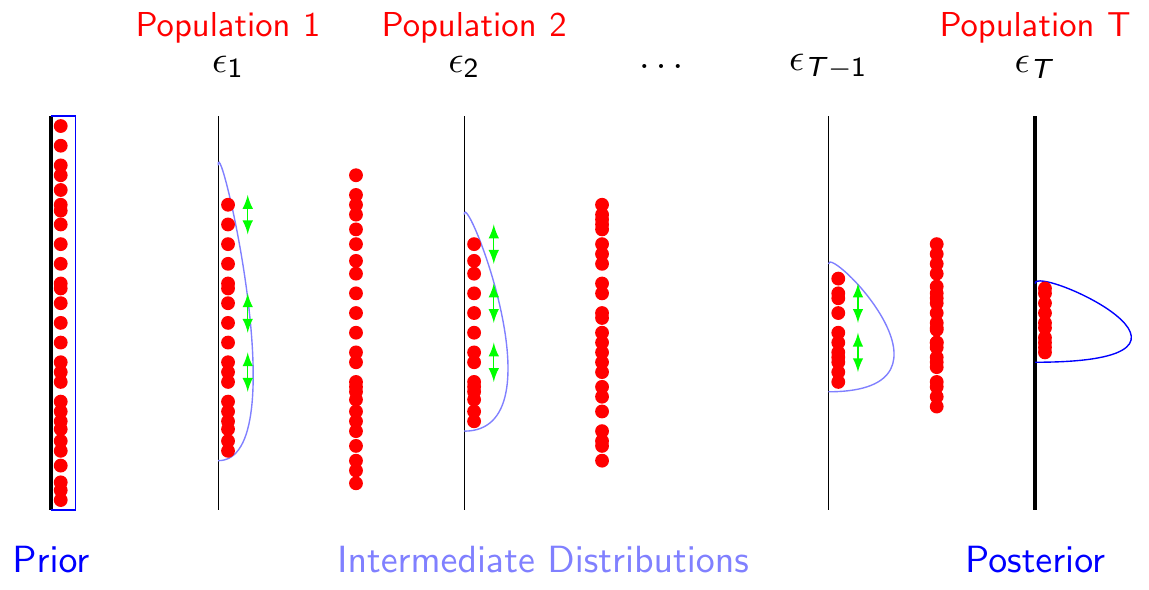}\label{fig:4smc}} 	
			\caption{\small{Schematic representation of ABC SMC.}}

\end{figure}
\small
\textbf{ABC SMC (Toni \textit{et al.}, 2009)}\\
\\
(a) As in ABC rejection, we define a prior distribution $P(\theta)$ and we would like to approximate a posterior distribution $P(\theta|D_0)$. In ABC SMC we do this sequentially by constructing intermediate distributions, which converge to the posterior distribution. We define a tolerance schedule $\epsilon_1 > \epsilon_2 > \ldots \epsilon_T \geq 0$.\\
\vspace{1.0cm}
\\
(b) We sample particles from a prior distribution until $N$ particles have been accepted (have reached the distance smaller than $\epsilon_1$). For all accepted particles we calculate weights (see \cite{Toni:2009p20998} for formulas and derivation). We call the sample of all accepted particles "Population 1".\\
\vspace{1.6cm}
\\
(c) We then sample a particle $\theta^{*}$ from population 1 and perturb it to obtain a perturbed particle $\theta^{**} \sim K(\theta|\theta^{*})$, where K is a perturbation kernel (for example a Gaussian random walk). We then simulate a dataset $D^{*} \sim f(D|\theta^{**})$ and accept the particle $\theta^{**}$ if $d(D_0,D^{**})\leq \epsilon_2$. We repeat this until we have accepted $N$ particles in population 2. We calculate weights for all accepted particles.\\
\vspace{1cm}
\\
(d) We repeat the same procedure for the following populations, until we have accepted $N$ particles of the last population $T$ and calculated their weights. Population $T$ is a sample of particles that approximates the posterior distribution.\\
\\
ABC SMC is computationally much more efficient than ABC rejection (see \cite{Toni:2009p20998} for comparison).

\newpage

\onecolumn 

\small
\textbf{ABC rejection for model selection\\}

\begin{figure}[thp]	

	\centering
	\subfigure[]{\includegraphics[width=4cm]{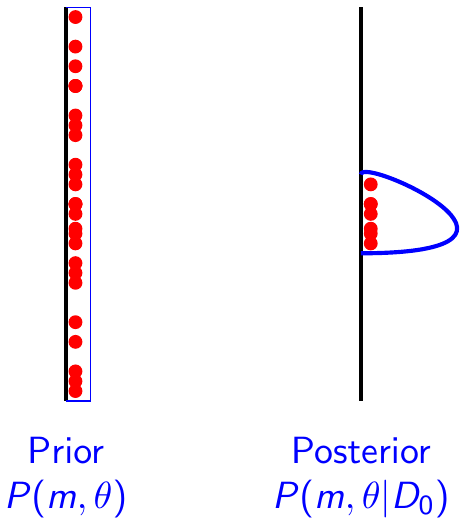}\label{fig:prior_posterior_msel}} \hspace{1cm}
	\subfigure[]{\includegraphics[width=9cm]{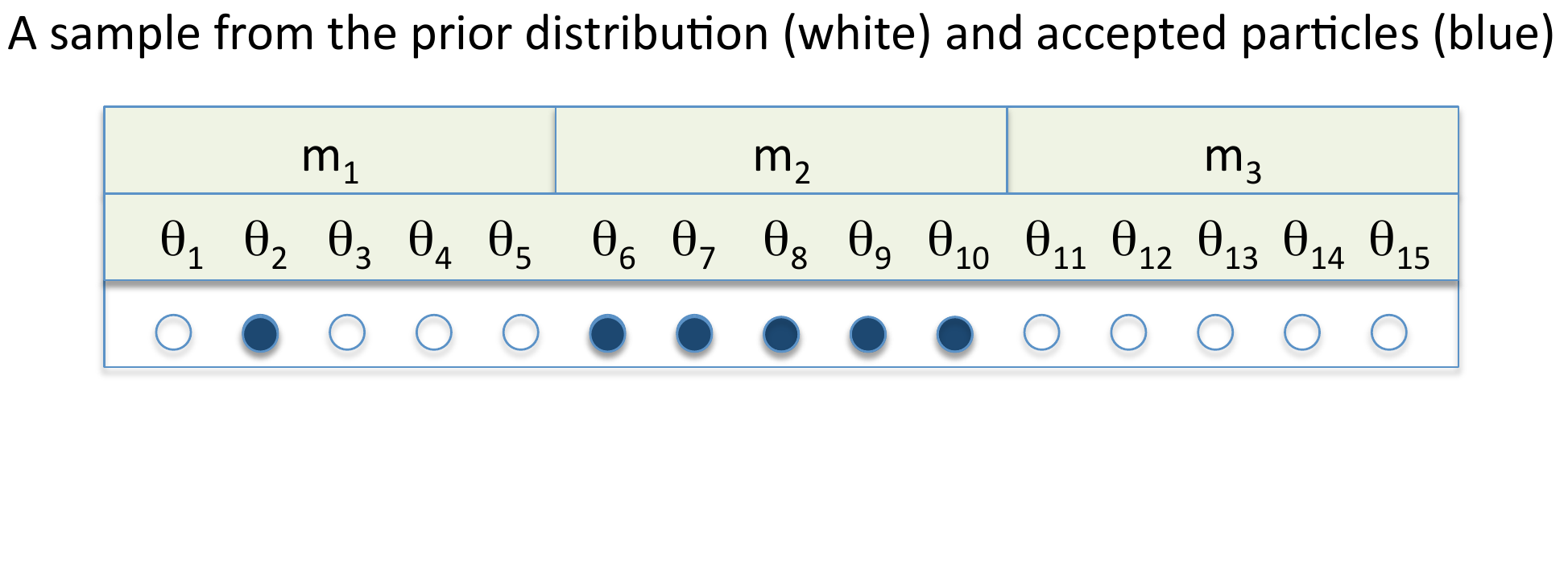}\label{fig:msel_rej}} \\
	\vspace{0.5cm}
	\subfigure[]{\includegraphics[width=5.0cm]{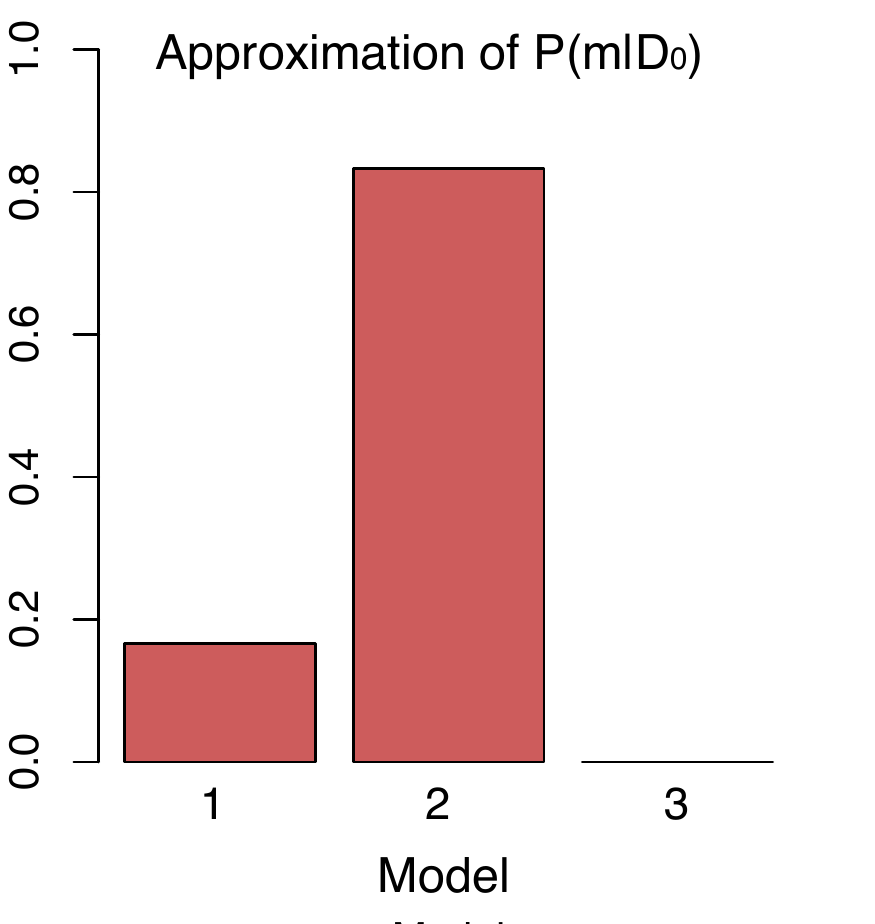}\label{fig:boxplot_rej_example}} \\		
	
			\caption{\small{(a) Prior and posterior distributions, $P(m,\theta)$ and $P(m,\theta|D_0)$, are now defined on a joint model and parameter space. (b) Particles $(m,\theta)$ are sampled from the prior distribution and accepted/rejected according to the distance between the simulated and experimental datasets. The accepted particles are shown in dark blue. (c) Six particles have been accepted: one from model 1 and five from model 2. The approximated marginal posterior probability of the model can be calculated as $P(m=1|D_0) = \frac{1}{6}$, $P(m=2|D_0) = \frac{5}{6}$, $P(m=3|D_0) = \frac{0}{6}$ . For illustrative purposes we have chosen a small number of particles. In principle this algorithm will yield consistent marginal posterior model distributions for $N \rightarrow \infty$.}}
\end{figure}

\newpage

\small
\textbf{ABC SMC for model selection\\}
\\

\begin{figure}[thp]	

	\centering
	\subfigure[]{\includegraphics[width=8cm]{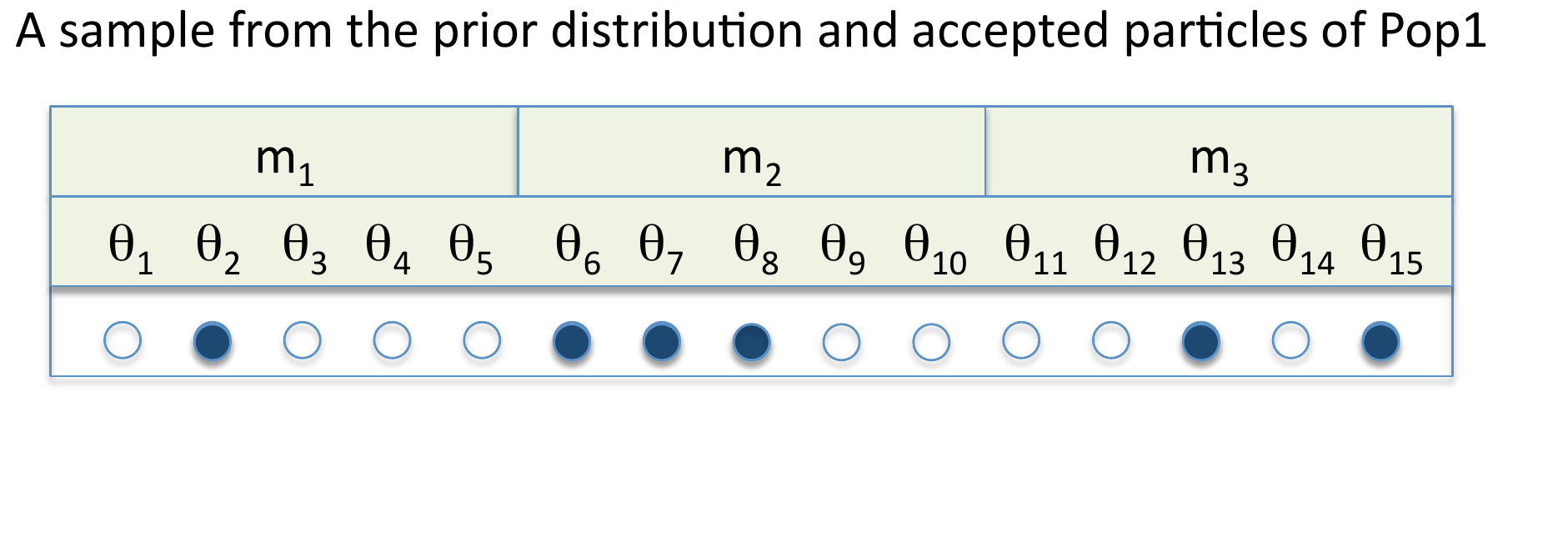}\label{fig:schematic_msel_pop1}} \hspace{0.2cm}
	\subfigure[]{\includegraphics[width=4.0cm]{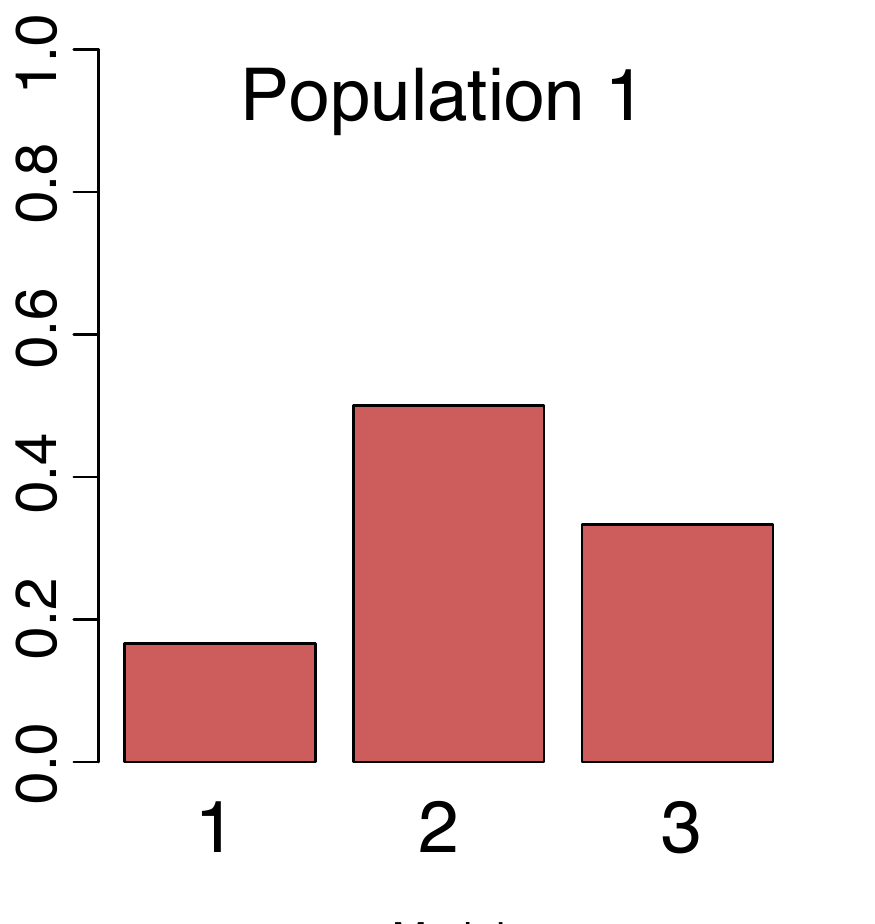}\label{fig:smc_populations_pop1}} \\	
	\subfigure[]{\includegraphics[width=6.5cm]{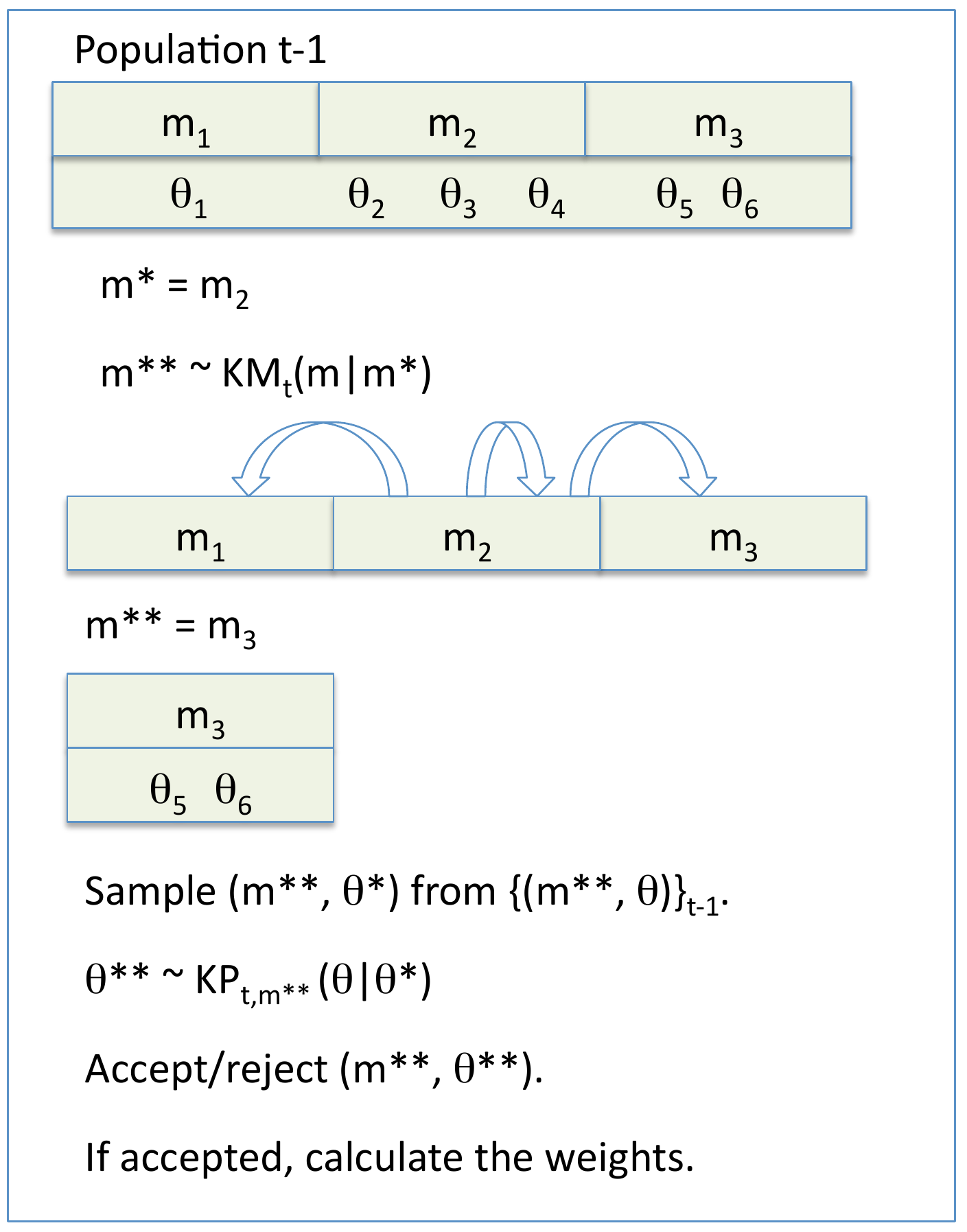}\label{fig:algorithm_schematic}} \hspace{0.1cm}
	\subfigure[]{\includegraphics[width=5.5cm]{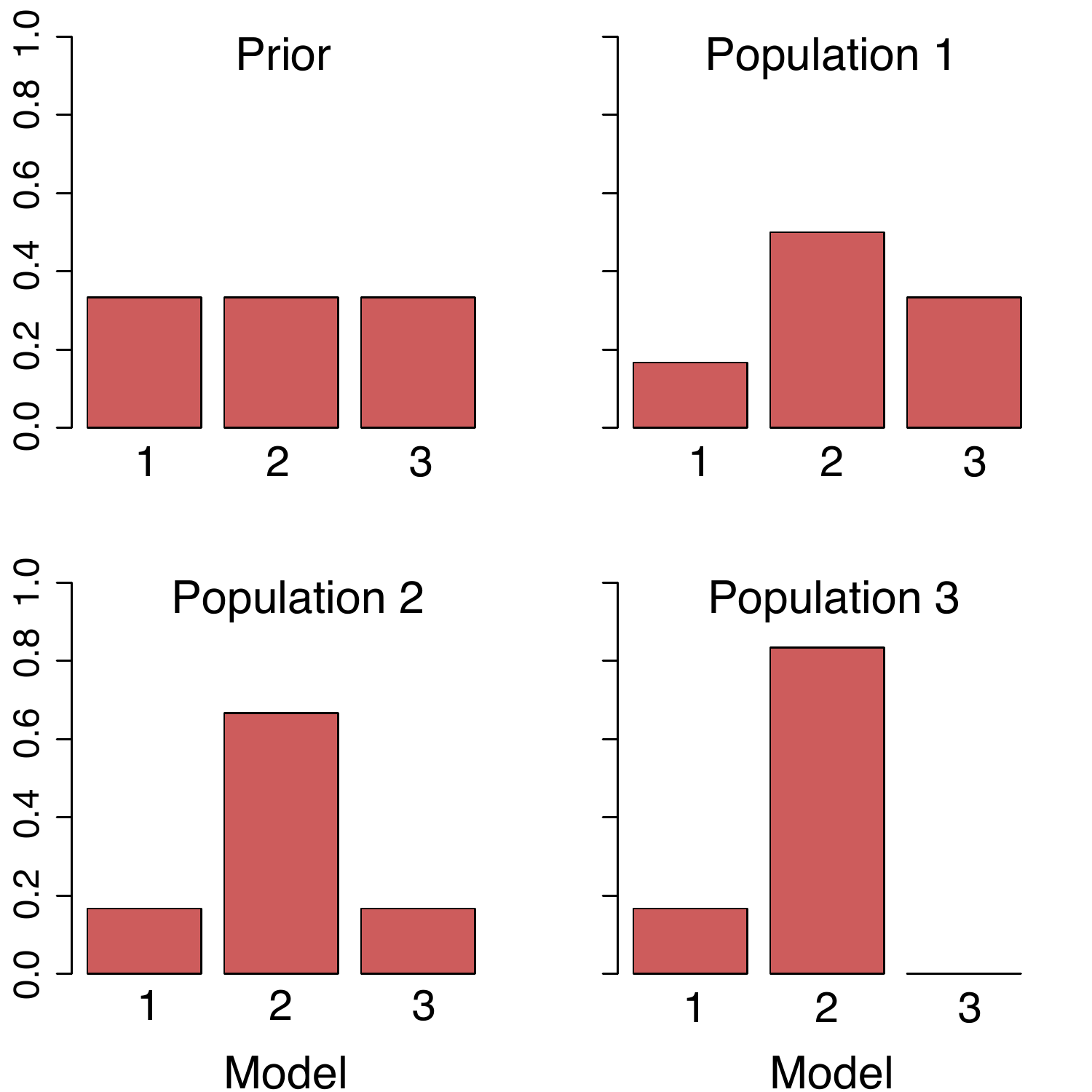}\label{fig:smc_populations_example}} 
\end{figure}

\newpage

\begin{figure}[thp]	

	\centering
			\caption{
			\small{
			\protect \\
			\protect \\
			(a) Particles are sampled from the prior $P(m,\theta)$ until $N$ particles have been accepted (in this example $N=6$ for illustration purposes, in practice $N$ should be larger), that is the distance is smaller than $\epsilon_1$. The weights $w_{t=1}$ are calculated for all accepted particles and normalized. \protect \\
			\protect \\
			(b) To obtain the marginal probability of $m$, we sum over all the weights corresponding to the model of interest: $P_{t=1}(m') = \sum_{m=m'}w_{t=1}(m,\theta)$. The histogram of the marginal population $1$ is presented. \protect \\
			\protect \\
			(c) This figure shows how we propose and accept a particle of population $t$, $t=2,\ldots,T$. We sample a model $m^{*}$ from population $t-1$ with probability $P_{t-1}(m^{*})$. For example, we might have sampled $m^{*} = m_2$. We then perturb the model using a model perturbation kernel $KM_t$ to obtain $m^{**} \sim KM_t(m|m^{*})$, for example $m^{**} = m_3$. After we have obtained the model $m^{**}$, we sample a parameter $\theta^{*}$ belonging to model $m^{**}$ from population $t-1$ and perturb it to obtain $\theta^{**} \sim KP_{t,m^{**}}(\theta|\theta^{*})$. We simulate a dataset $D^{*}$ for a particle $(m^{**},\theta^{**})$ and accept ($d(D_0,D^{*})\leq \epsilon_t$) or reject the particle. If a particle is accepted, we calculate its weight $w_t(m^{**},\theta^{**})$. \protect \\
			\protect \\
			(d) When $N$ particles of population $t$ have been accepted, we normalize the weights and marginalize them in order to obtain the marginal intermediate population of the model $P_t(m)$. We continue until population $T$, which is the approximation of the joint posterior distribution $P(m,\theta|D_0)$. The quantities of interest are the intermediate and the last marginal population of the model. In this example there are three populations, $T=3$.
			}
			}
			
\end{figure}

\newpage

\footnotesize


\end{document}